\documentclass{eas}
\usepackage{graphicx}
\begin{document}

\title{$<$3D$>$ NLTE line formation in the atmospheres of red supergiants}
\author{Maria Bergemann}\address{Max-Planck-Institute for Astrophysics, Karl-Schwarzschild-Str.1, D-85741 Garching, Germany}
\author{Rolf P. Kudritzki}\address{Institute for Astronomy, University of Hawaii, 2680 Woodlawn Drive, Honolulu, HI 96822}
\author{Ben Davies}\address{Institute of Astronomy, Univerity of Cambridge, Madingley Road, Cambridge, CB3 OHA, UK}
\author{Bertrand Plez}\address{Laboratoire Univers et Particules de Montpellier, Universit\'e Montpellier 2, CNRS, F-34095 Montpellier, France}
\author{Zach Gazak}\sameaddress{2}
\author{Andrea Chiavassa}\address{Laboratoire Lagrange, UMR7293, Universit\'e de Nice Sophia-Antipolis, CNRS, Observatoire de la Cote dAzur, BP 4229, F-06304 Nice cedex 4, France}
\begin{abstract}
Red supergiants with their enormous brightness at J-band are ideal probes of cosmic chemical composition. It is therefore crucial to have realistic models of radiative transfer in their atmospheres, which will permit determination of abundances accurate to 0.15 dex, the precision  attainable with future telescope facilities in galaxies as distant as tens of Mpc. Here, we study the effects of non-local thermodynamic equilibrium (NLTE) on the formation of iron, titanium, and silicon lines, which dominate J-band spectra of red supergiants. It is shown that the NLTE radiative transfer models enable accurate derivation of metallicity and effective temperature in the J-band.  We also discuss consequences for RSG spectrum synthesis in different spectral windows, including the heavily TiO-blanketed optical region, and atmospheric structure. We then touch upon challenges of NLTE integration with new generation of 3D hydrodynamical RSG models and present the first calculations of NLTE spectra with the mean 3D model of Betelgeuse.
\end{abstract}
\maketitle
\runningtitle{$<$3D$>$ NLTE line formation for red supergiants}
%
\section{Introduction}
Chemical composition is one of the key observable characteristics of star forming galaxies in the nearby and in the high redshift universe. So far, most of our information about their metal content has been obtained from the analysis of strong emission lines from H II regions. However, measurements of galaxy metallicities
are then uncertain by a large factor because of the systematic uncertainties inherent to this 'strong-line' method. Furthermore, the method yields basically only the oxygen abundance, which is then taken as a placeholder for the overall metallicity. In this case, there is no information on abundance ratios, which can be a powerful diagnostic of the chemical enrichment history.

An alternative approach avoiding these weaknesses is the spectroscopic analysis of supergiant stars.  Here, much progress has been made through the optical spectroscopy of blue supergiants in the Milky Way and few other galaxies of the Local group. For
extragalactic astrophysics, however, red supergiants (RSGs) are more
promising candidates. The spectral energy distribution of RSG's peaks in the
infrared, where interstellar extinction is reduced. Particularly attractive for 
quantitative spectroscopy is the J-band, which contains many isolated atomic
lines. Spatial resolution in the infrared is also higher than in the optical 
thanks to more efficient corrections by adaptive optics systems. Extragalactic RSGs are therefore ideal targets for spectroscopy with future telescope facilities, such as the European Extremely Large Telescope. The abundances of various chemical elements could be then directly measured out to distances of 70 Mpc, far beyond our Local Group of galaxies.

The analysis of RSG spectra, however, is a challenging task. The major complexities arise due to very low gravities, stipulating departures from Local Thermodynamic Equilibrium (LTE) in their photospheres, and violent surface convection. 

We present a detailed investigation of non-LTE (NLTE) line formation in RSG atmospheres using new atomic models of Fe, Si, and Ti computed with state-of-the-art atomic data. The radiative transfer calculations are performed with 1D hydrostatic and mean 3D hydrodynamic model atmospheres. The effects on spectrum synthesis and element abundance determinations from the atomic lines important in the near-IR spectra of RSG's are quantified.

\section{Line formation and atomic models}

\begin{figure}
\includegraphics[width=8cm,angle=90]{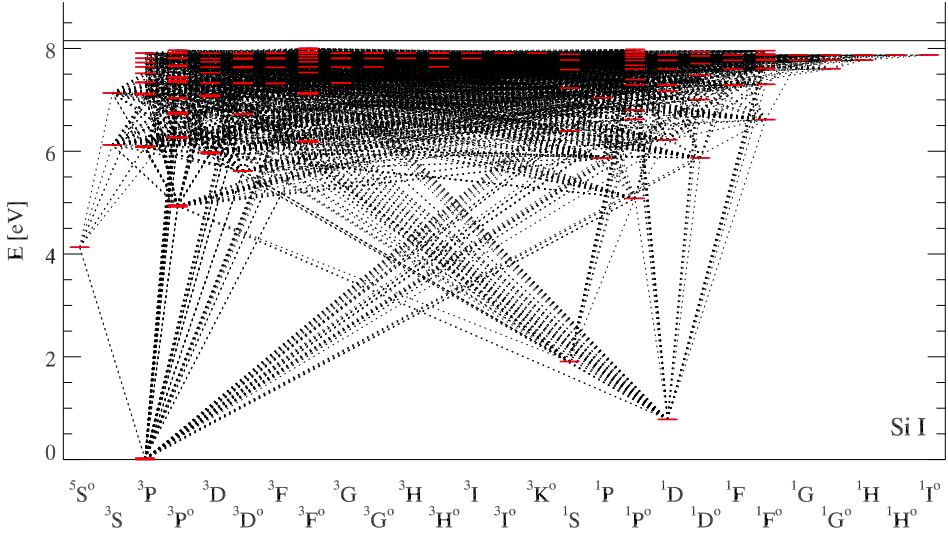}
\caption{The Grotrian diagram of the Si I model atom.}
\label{fig1}
\end{figure}

The atomic models of Ti, Fe, and Si have been described elsewhere (Bergemann et al. \cite{B12,B13}). In short, they were constructed from the best available atomic data, including the experimental data from NIST, theoretically-predicted energy levels and radiative transitions from the Kurucz's database, and photoionization cross-sections from TOPbase. The model atoms were tested by performing test calculations for several benchmark stars with well-known physical parameters, including the Sun and Arcturus. Fig. \ref{fig1} shows the Si I atom.

\begin{figure}
\includegraphics[width=8cm,angle=-90]{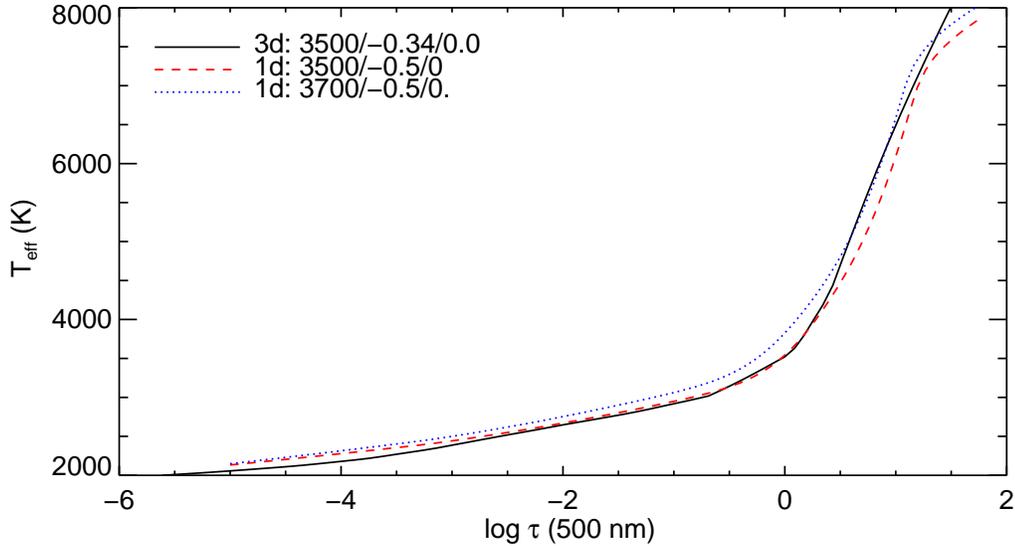}
\caption{The 1D MARCS and $<$3D$>$ model atmospheres used for NLTE radiative transfer. Stellar parameters ($T_{\rm eff}$, $\log g$, and [Fe/H]) are indicated in the plot.}
\label{fig2}
\end{figure}

As a basis for NLTE statistical equilibrium calculations, we use two types of model atmospheres. The MARCS model atmospheres (Gustafsson et al. 2008) are 1D hydrostatic models, widely used in spectroscopy of cool stars. We also attempt the non-LTE line formation with the spatially-averaged models constructed from the CO$^{5}$BOLD star-in-a-box simulations of stellar surface convection (see below). The  model tested in this work is that from Chiavassa et al. (\cite{Chia11}), \texttt{st35gm03n13}, with luminosity $89000 L_{\rm Sun}$ and $T_{\rm eff} = 3430$ K, and the $<$3D$>$ model is computed by a spherical averaging of the original simulation (Fig. \ref{fig2}). The details of the averaging procedure will be provided in a forthcoming publication. This is the first time that such models are coupled to NLTE radiative transfer with the goal to understand the impact on the RSG spectroscopic analysis. The DETAIL code is used for the calculation of NLTE occupation numbers. 

Line profiles are computed with the code SIU using the level departure coefficients from DETAIL. SIU and DETAIL share the same physics and line lists. For all details such as model structure and geometry, background opacities, solar abundance mixture, etc. we refer the reader to Bergemann et al. (\cite{B12,B13}).

\section{Abundances from non-LTE and LTE models}

In what follows, we first discuss the results obtained with the MARCS 1D hydrostatic model atmospheres because a large grid of such models for RSG parameters is available and systematic vartiations of NLTE effects with $T_{\rm eff}$, $\log g$, and [Fe/H] can be well-understood. Given the very complex structure of the atoms, we focus only on the energy levels and transitions related to the near-IR J-band analysis of RSG stars (Si I: multiplet 4s$^3$P$^{\rm o}$ - 4p$^3$D, Ti I: b$^3$F - z$^3$D$^{\rm o}$)

The inclusion of NLTE effects changes the titanium abundances dramatically compared to LTE and with different abundance correction factors $\Delta$ for different temperatures and metallicities. The NLTE abundance corrections\footnote{The NLTE correction is a logarithmic correction, which has to be applied to an LTE abundance determination of a specific line to obtain the correct value corresponding to the use of NLTE line formation. We calculate these corrections at each point of our model grid for each line by matching the NLTE equivalent width through varying the Si abundance in the LTE calculations.} vary smoothly between $-0.4$ dex and $+0.2$ dex as a function of effective temperature, which is caused by the interplay of continuum processes and line transition cascades to and from the relevant atomic energy levels. The NLTE effects on the J-band Fe I lines are much smaller. The abundance corrections become non-negligible,  $\Delta \geq 0.1$ dex, only for $T_{\rm eff} > 4000$ K, when radiative pumping to the higher levels takes over. 

In what concerns the J-band Si I lines, the situation resembles the classical case of the resonance line formation: photo-absorption in the line transition and competing de-population of the upper levels by spontaneous emission and collisional de-excitation. The NLTE abundance corrections are substantial and are always negative, with values between $-0.4$ to $-0.1$ dex. In contrast to Ti I and Fe I, the NLTE effects in the Si I J-band lines are weaker at higher effective temperature.

\section{Non-LTE and mean 3D RHD model atmospheres}

At present, it is not possible to systematically evaluate the effect of the 1D hydrostatic equilibrium approximation on the determination of abundances from the observed spectra of red supergiants. For such calculations, full grids of 3D RHD models in the relevant space of stellar parameters are necessary, which are not yet available.

The only realistic non-gray 3D radiative hydrodynamics simulation of convection for a single RSG model was performed by Chiavassa et al. (\cite{Chia11}). They demonstrated that there is no unique 1D static model which would provide similar observable characteristics, i.e., SED shape and colors, as the analogous 3D RHD model. To match the optical spectrum dominated by TiO opacity, a $\sim 100$ K cooler 1D model is needed, whereas the IR spectral window can be best matched using a $\sim 200$ K hotter 1D model. The NLTE effects, however, are sensitive to the radiation field across the whole spectrum, from UV to the IR. Thus, it is not possible to predict the changes in the NLTE abundance corrections simply by using a cooler or hotter 1D hydrostatic model. 

\begin{figure}
\includegraphics[width=8cm,angle=-90]{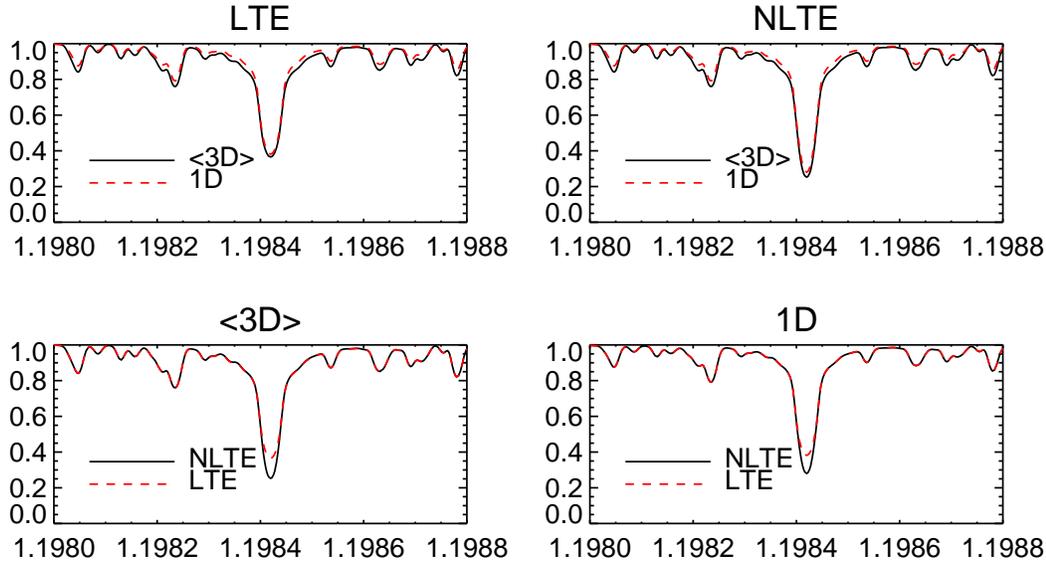}
\caption{NLTE line formation with the 1D static and $<$3D$>$ model atmosphere for the Si I line at 4.8 eV.}
\label{fig3}
\end{figure}

\begin{figure}
\includegraphics[width=8cm,angle=-90]{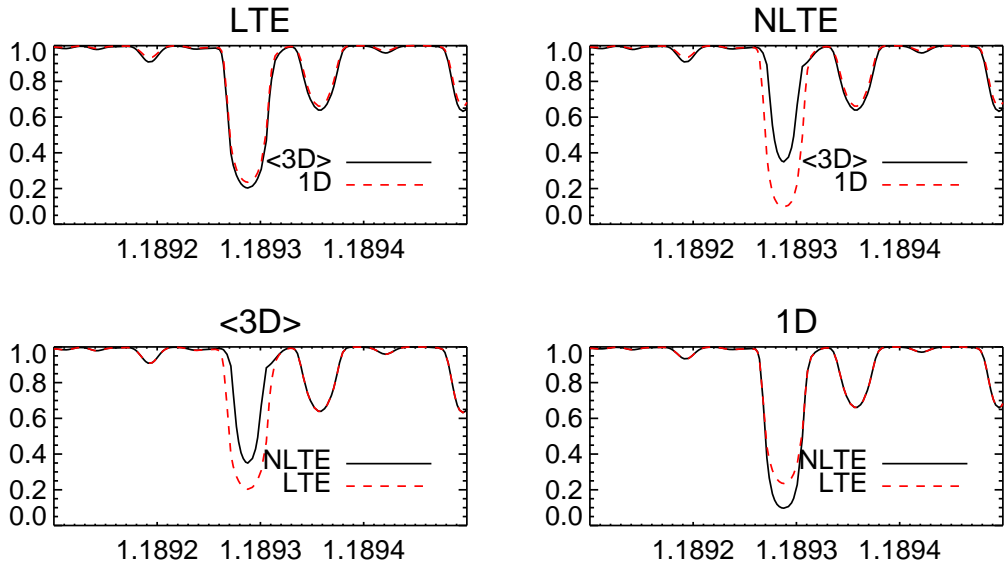}
\caption{NLTE line formation with the 1D static and $<$3D$>$ model atmosphere for the Ti I line at 1.4 eV.}
\label{fig4}
\end{figure}

We have performed full NLTE spectrum synthesis with the $<$3D$>$ model atmosphere of a cool RSG star (see Sec. 2). The model has $T_{\rm eff} = 3430$ K, $\log g = -0.34$, and [Fe/H] $=0$. For comparison, we used a MARCS 1D static model computed for the same physical parameters albeit a lower \textit{effective} surface gravity, $\log g = -0.5$, comparable to what is necessary to account for the effects of turbulent pressure in the static models (Chiavassa et al. \cite{Chia11}). As expected, the differences between the $<$3D$>$ and 1D static results depend critically on the type of NLTE effects participating in the distribution of atomic level populations. This is illustrated in Fig. \ref{fig3} and Fig. \ref{fig4} for the lines of Si I and Ti I, respectively. 

Si I is a majority ion because of its very high first ionization potential, $E \sim  8.15$ eV. The NLTE effects in the high-excitation (4.9 eV) Si I line are entirely due to the radiation field in the line itself, and variation of the background radiation field emerging from the deep atmosphere has no effect. The line formation takes place in the optical depth range $-2 < \log \tau_{\rm 500 nm} < 0$, where the thermal and pressure structures of the 1D and $<$3D$>$ models are identical. As a consequence, the Si I line profiles computed with the MARCS and the $<$3D$>$ are very similar and both become notably darker in the cores compared to LTE. This implies negative NLTE abundance corrections in either case.

In contrast, for the near-IR low-excitation (1.4 eV) Ti I lines the effect is different. Ti I is a minority ionization stage (N$_{\rm Ti I}/N_{\rm Ti II} \sim 10^{-2}$) at all optical depths. Thus, photoionization and pumping by the radiation field originating from the atmospheric depths, where the $<$3D$>$ models is $\sim 200$ K warmer than the 1D model, is amplified. This causes the NLTE line weakening so that an LTE analysis with 1D or $<$3D$>$ models will strongly overestimate the abundance.

We also performed LTE radiative transfer calculations with the $<$3D$>$ models in the optical region centered on the TiO molecular bands. The lines form at intermediate optical depths and, given the similarity of the mean model structures, no significant differences were found between both models in LTE. This result is not unexpected. Molecules are very sensitive to temperature and pressure inhomogeneities 
(Uitenbroek \& Criscuoli \cite{Uit11}), thus the correct physical picture of TiO line formation demands NLTE calculations with full 3D RHD simulations.

\section{Conclusions}

We investigated the effects of non-local thermodynamic equilibrium on the formation of strong Fe I, Ti I, and Si I lines, which are very prominent in the J-band spectra of red supergiants (RSG). Since our long-term goal is a comprehensive near-IR analysis of RSG's, only the model atmospheres with $T_{\rm eff} \leq 4400$ K and $\log g \leq 1$ are considered.

LTE models are fully inadequate for spectral analysis of red supergiants, both with 1D hydrostatic and 3D hydrodynamic models. The NLTE radiative transfer models lead to significantly different line profiles compared to LTE. These differences are most pronounced if the averaged 3D RHD models are used, which is also known from the $<$3D$>$ NLTE analysis of warmer FGK stars (Bergemann et al. \cite{B12b}). The synthetic NLTE line profiles and abundances inferred using these are similar in 1D and $<$3D$>$ for the high-excitation Si I lines, but they are radically different for the low-excitation Ti I lines. It is thus recommended, in order not to introduce systematic effects in determinations of RSG chemical composition, to avoid low-excitation lines of neutral metals in the near-IR.

LTE spectrum synthesis in the optical spectral window, centered on the heavily TiO-blanketed region, did not reveal any significant differences between using the 1D static and the $<$3D$>$ model. NLTE calculations with full 3D RHD simulations of RSG atmospheres are necessary in this case.


\end{document}